\documentclass{elsart}
\usepackage{graphics}
\usepackage{epsfig}
\usepackage{color}
\usepackage{amsmath}
\usepackage{amssymb}

\definecolor{mygreen}{rgb}{0,.8,0}

\def\ifm#1{\relax\ifmmode#1\else$#1$\fi}
  
\def\f{\ifm{\phi}}  \def\ff{\f--factory} \def\epm{\ifm{e^+e^-}}
  \def\x{\ifm{\times}}
  
\def\pt#1,#2,{\ifm{#1\x10^{#2}}}  
\renewcommand{\to}{\ensuremath{\rightarrow}}

\makeatletter
\newdimen\z@ \z@=0pt 
\newskip\z@skip \z@skip=0pt plus0pt minus0pt
\def\m@th{\mathsurround=\z@}
\def\ialign{\everycr{}\tabskip\z@skip\halign} 
\def\eqalign#1{\null\,\vcenter{\openup\jot\m@th
  \ialign{\strut\hfil$\displaystyle{##}$&$\displaystyle{{}##}$\hfil
      \crcr#1\crcr}}\,}
\makeatother
\newcommand{\affuni}[2]{Dipartimento di Fisica dell'Universit\`a #1, #2, Italy.}

\newcommand{\affinfnm}[2]{INFN Sezione di #2, #2, Italy.}
\newcommand{\affinfnn}[2]{INFN Sezione di #1, #2, Italy.}

\newcommand{\kl}{\mbox{$K_L$}}
\newcommand{\ks}{\mbox{$K_S$}}

\newcommand{\Pphi}{\ensuremath{\phi}}

\newcommand{\eV}{{e\kern-.07em V}}
\newcommand{\kloe}{K{\kern-.07em LOE} }
\newcommand{\dafne}{D{\kern-.07em A$\Phi$NE}}
\newcommand{\keV}{{\rm \,k\eV}}
\newcommand{\MeV}{{\rm \,M\eV}}
\newcommand{\GeV}{{\rm G\eV}}
\newcommand{\ps}{{\rm \,ps}}
\newcommand{\ns}{{\rm \,ns}}

\newcommand{\cm}{{\rm \,cm}}
\newcommand{\m}{{\rm \,m}}

\newcommand{\rad}{{\rm \,rad}}

\newcommand{\T}{{\rm \,T}}
\newcommand{\Lnb}{\ensuremath{\rm \, nb^{-1}}}

\newcommand{\Lfb}{\ensuremath{\rm \, fb^{-1}}}

\newcommand{\DKSpippim}{\ensuremath{K_S\rightarrow\pi^+\pi^-}}
\newcommand{\pippim}{\ensuremath{\pi^+\pi^-}}
\newcommand{\pipi}{\ensuremath{\pi\pi}}
\newcommand{\pimu}{\ensuremath{\pi\mu}}
\newcommand{\DKSpiopio}{\ensuremath{K_S\rightarrow\pi^0\pi^0}}
\newcommand{\DKSee}{\ensuremath{K_S\rightarrow e^+e^-}}

\newcommand{\DKSgg}{\ensuremath{K_S\rightarrow \gamma \gamma}}

\newcommand{\Dphipippimpio}{\ensuremath{\phi\rightarrow\pi^+\pi^-\pi^0}}

\newcommand{\trepi}{\ensuremath{3\pi}}
\newcommand{\UL}{\ensuremath{\mathrm{UL}}}
\newcommand{\BR}[1]{\ensuremath{\mathrm{BR}(#1)}}

\newcommand{\kcr}{\ensuremath{K_\mathrm{crash}}}

\newcommand{\koko}{\ensuremath{K^0\bar{K}^0}}

\newcommand{\ee}{\ensuremath{e^+e^-}}
\newcommand{\mumu}{\ensuremath{\mu^+\mu^-}}
\newcommand{\Mee}{\ensuremath{M_{ee}}}
\newcommand{\pmiss}{\ensuremath{p_{\mathrm{miss}}}}

\begin{document}

\begin{frontmatter}
\title{Search for the $\DKSee$ decay with the \kloe\ detector}
\collab{The KLOE Collaboration}

\author[Na,infnNa]{F.~Ambrosino},
\author[Frascati]{A.~Antonelli},
\author[Frascati]{M.~Antonelli},
\author[Roma2,infnRoma2]{F.~Archilli},
\author[Karlsruhe]{P.~Beltrame},
\author[Frascati]{G.~Bencivenni},
\author[Frascati]{S.~Bertolucci},
\author[Roma1,infnRoma1]{C.~Bini},
\author[Frascati]{C.~Bloise},
\author[Roma3,infnRoma3]{S.~Bocchetta},
\author[Frascati]{F.~Bossi},
\author[infnRoma3]{P.~Branchini},
\author[Frascati]{P.~Campana},
\author[Frascati]{G.~Capon},
\author[Roma3,infnRoma3]{D.~Capriotti},
\author[Frascati]{T.~Capussela},
\author[Roma3,infnRoma3]{F.~Ceradini},
\author[Frascati]{P.~Ciambrone},
\author[Roma1]{F.~Crucianelli},
\author[Frascati]{E.~De~Lucia},
\author[Roma1,infnRoma1]{A.~De~Santis},
\author[Frascati]{P.~De~Simone},
\author[Roma1,infnRoma1]{G.~De~Zorzi},
\author[Karlsruhe]{A.~Denig},
\author[Roma1,infnRoma1]{A.~Di~Domenico},
\author[infnNa]{C.~Di~Donato},
\author[Roma3,infnRoma3]{B.~Di~Micco},
\author[Frascati]{M.~Dreucci},
\author[Frascati]{G.~Felici},
\author[Frascati]{M.~L.~Ferrer},
\author[Roma1,infnRoma1]{S.~Fiore},
\author[Roma1,infnRoma1]{P.~Franzini},
\author[Frascati]{C.~Gatti},
\author[Roma1,infnRoma1]{P.~Gauzzi},
\author[Frascati]{S.~Giovannella},
\author[infnRoma3]{E.~Graziani},
\author[Karlsruhe]{W.~Kluge},
\author[Moscow]{V.~Kulikov},
\author[Frascati]{G.~Lanfranchi},
\author[Frascati,StonyBrook]{J.~Lee-Franzini},
\author[Karlsruhe]{D.~Leone},
\author[Frascati,Energ]{M.~Martini},
\author[Na,infnNa]{P.~Massarotti},
\author[Na,infnNa]{S.~Meola},
\author[Frascati]{S.~Miscetti},
\author[Frascati]{M.~Moulson},
\author[Frascati]{S.~M\"uller},
\author[Frascati]{F.~Murtas},
\author[Na,infnNa]{M.~Napolitano},
\author[Roma3,infnRoma3]{F.~Nguyen},
\author[Frascati]{M.~Palutan},
\author[infnRoma1]{E.~Pasqualucci},
\author[infnRoma3]{A.~Passeri},
\author[Frascati,Energ]{V.~Patera},
\author[Na,infnNa]{F.~Perfetto},
\author[Frascati]{P.~Santangelo},
\author[Frascati]{B.~Sciascia},
\author[Frascati,Energ]{A.~Sciubba},
\author[Frascati]{A.~Sibidanov},
\author[Frascati]{T.~Spadaro},
\author[Roma1,infnRoma1]{M.~Testa},
\author[infnRoma3]{L.~Tortora},
\author[infnRoma1]{P.~Valente},
\author[Frascati]{G.~Venanzoni},
\author[Frascati,Energ]{R.Versaci}

\address[Frascati]{Laboratori Nazionali di Frascati dell'INFN, 
Frascati, Italy.}
\address[Karlsruhe]{Institut f\"ur Experimentelle Kernphysik, 
Universit\"at Karlsruhe, Germany.}
\address[Na]{Dipartimento di Scienze Fisiche dell'Universit\`a 
``Federico II'', Napoli, Italy}
\address[infnNa]{INFN Sezione di Napoli, Napoli, Italy}
\address[Energ]{Dipartimento di Energetica dell'Universit\`a 
``La Sapienza'', Roma, Italy.}
\address[Roma1]{\affuni{``La Sapienza''}{Roma}}
\address[infnRoma1]{\affinfnm{``La Sapienza''}{Roma}}
\address[Roma2]{\affuni{``Tor Vergata''}{Roma}}
\address[infnRoma2]{\affinfnn{Roma Tor Vergata}{Roma}}
\address[Roma3]{\affuni{``Roma Tre''}{Roma}}
\address[infnRoma3]{\affinfnn{Roma Tre}{Roma}}
\address[StonyBrook]{Physics Department, State University of New 
York at Stony Brook, USA.}
\address[Moscow]{Institute for Theoretical 
and Experimental Physics, Moscow, Russia.}

\corauth[cor1]{Corresponding author: flavio.archilli@lnf.infn.it}
\corauth[cor2]{Corresponding author: tommaso.spadaro@lnf.infn.it}

\begin{abstract}
 We present the result of a direct search for the decay \DKSee, obtained with
a sample of $\epm \to \Pphi \to \ks\kl$ events produced at \dafne, the Frascati \ff, for an integrated luminosity
of $1.9\Lfb$.
The search has been performed using a pure \ks\ beam tagged by the simultaneous detection of a 
\kl\ interaction in the calorimeter. Background rejection has been optimized by using both kinematic and
particle identification cuts. We find $\BR\DKSee < 9\times10^{-9}$ at 90\%~CL, 
which improves by an order of magnitude on the previous best limit.

\end{abstract}
\end{frontmatter}

\section{Introduction}
\begin{figure}[h!]
  \begin{center}    
    \includegraphics[totalheight=2.5cm]{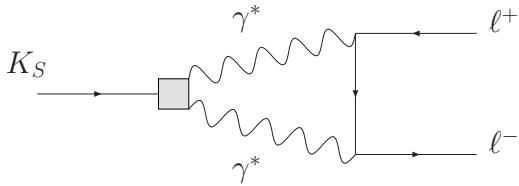}
    \caption{Long distance contribution to $\ks \rightarrow \ell^+ \ell^-$ process, mediated by two-photon exchange.}
    \label{fig:sec1.fig2}
  \end{center}
\end{figure}

The \kl, \ks\ decays into leptons pairs (\ee, \mumu) are due to $\Delta S = 1$ flavour-changing neutral-currents (FCNC) transitions.
The decay amplitudes receive contributions both from long distance (LD) effects, dominated by the $2\gamma$ 
intermediate state shown in Fig.~\ref{fig:sec1.fig2}, and from short-distance (SD) effects, due to box
and penguin diagrams via $W$,$Z$ exchange.
The SD contribution can be rather precisely evaluated in the Standard Model (SM) 
so that a theoretical evaluation of the LD contribution would allow a comparison
of experimental results against predictions that may represent
a significant test of the SM.
For the \kl\ decay the evaluation of the LD contribution needs a model for the $\kl \to \gamma^* \gamma^*$
form factor, while for the \ks\ it can be determined at lowest order in the chiral perturbation theory. In this case one obtains~\cite{kseetheory}:
\begin{equation}
\begin{array}{rcl}
\Gamma(K_S \rightarrow \mumu)/\Gamma(K_S \rightarrow \gamma \gamma) &\simeq& 2 \times 10^{-6},\\
\Gamma(K_S \rightarrow \ee)/\Gamma(K_S \rightarrow \gamma \gamma) &\simeq& 8 \times 10^{-9}.
\end{array}
\end{equation}

Using the present average~\cite{pdg06} for BR(\DKSgg), we evaluate 
$\BR\DKSee \simeq 2\times 10^{-14}$.
A value significantly higher would point to new physics.
The best experimental limit for BR(\DKSee) has been obtained
by CPLEAR~\cite{cplearksee}, and it is equal to $1.4\times 10^{-7}$, at $90\%$~CL.
Here we present a new search for this decay, which improves on the previous limit
by more than an order of magnitude.

\section{Experimental setup}
\label{sec:expsetup}

The data were collected with the \kloe\ detector at \dafne, the Frascati \ff. 
\dafne\ is an \epm\ collider that operates at a 
center-of-mass energy of $\sim 1020\MeV$, the mass of the \Pphi\ meson.
Positron and electron beams of equal energy collide at an angle 
of $(\pi -0.025) \rad$, producing \Pphi\ mesons with a small momentum in the
horizontal plane: $p_\phi \approx 13\MeV /c$. \Pphi\ mesons decay 
$\sim 34\%$ of the time into nearly collinear \koko\ pairs. 
Because $J^{PC}(\phi)=1^{--}$, the kaon pair is in an antisymmetric state, 
so that the final state is always \ks\kl. The contamination from \kl\kl\ 
and \ks\ks\ final states is negligible.
Therefore, the detection of a \kl\ signals the presence of a \ks\ of known
momentum and direction, independently of its decay mode. 
This technique is called \ks\ {\it tagging}. 
The sample analyzed corresponds to an integrated luminosity of $\sim 1.9\Lfb$,
 yielding $\sim 2$ billion \ks\kl\ pairs. 
\begin{figure}[h!]
  \begin{center}    
    \includegraphics[totalheight=8.cm]{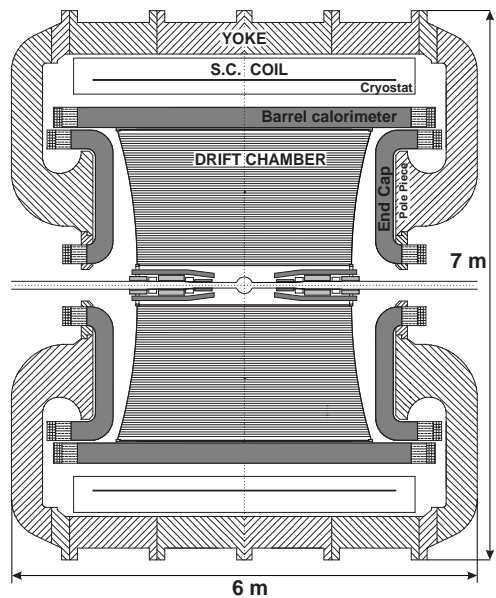}
    \caption{Vertical cross section of the KLOE detector.}
    \label{fig:kloedetector}
  \end{center}
\end{figure}

The \kloe\ detector (Fig.~\ref{fig:kloedetector}) consists of a 
large cylindrical drift chamber (DC), surrounded by a lead/scintillating-fiber
sampling calorimeter (EMC). A superconducting coil surrounding the calorimeter
provides a $0.52\T$ magnetic field. The drift chamber~\cite{drift} is $4\m$
in diameter and $3.3\m$ long. The chamber shell is made of carbon-fiber/epoxy
composite, and the gas used is a $90\%$ helium, $10\%$ isobutane mixture.
These features maximize transparency to photons and reduce \kl$\to$\ks\
regeneration and multiple scattering. 
The momentum resolution is $\sigma(p_\perp)/p_\perp = 0.4\%$, 
and the \DKSpippim\ invariant mass is reconstructed with a resolution of $\sim1\MeV$.

The calorimeter~\cite{calo} is divided into a barrel and two endcaps, 
covering $\sim98\%$ of the solid angle.
The modules are read out at both ends by photomultiplier tubes. The arrival
times of particles and the three-dimensional positions of the energy deposits
are determined from the signals at the two ends. The read-out granularity
is $\sim 4.4\times 4.4\, \cm ^2$; fired ``cells'' close in space and time
are grouped into a 
``calorimeter cluster''. For each cluster, the energy $E_{cl}$ is the
sum of the cell energies, and the time $t_{cl}$ and the position 
$\mathrm{r}_{cl}$ are calculated as energy-weighted averages over the fired 
cells. The energy and time resolutions are $\sigma_E/E = 5.7\%/ \sqrt{E(\GeV)}$
and $\sigma_t=57\ps / \sqrt{E(\GeV)}\oplus 100\ps$, respectively.

The trigger~\cite{ctrig} used for this analysis requires two local energy deposits 
above a threshold of $50\MeV$ in the barrel and $150\MeV$ in the endcaps. 
The trigger has a large time spread with respect to the beam crossing
time. However, it is synchronized with the machine RF divided by 4, 
$T_{\mathrm sync}\sim 10.8\ns$, with an accuracy of $50\ps$.
An estimate of the event production time ($T_0$) is 
determined offline. 

The response of the detector to the decays of interest and the various
backgrounds were studied by using the \kloe\ Monte Carlo (MC) simulation
program~\cite{filfo}. Changes in the machine operation and background conditions
are simulated on a run-by-run basis.
The beam energies and the crossing angle are obtained from the analysis of 
Bhabha scattering events with $e^\pm$ polar angle above 45 degrees.
The average value of the center-of-mass energy is evaluated with a precision
of $30\keV$ for each $100\Lnb$ of integrated luminosity. 

To study the background rejection, a MC sample of \Pphi\ decays to all possible 
final states has been used, equivalent to an integrated luminosity of $\sim 2.1\Lfb$.
A MC sample of $45\,000$ \DKSee\ events has been also produced, corresponding to a BR of $1.6\times 10^{-4}$.
This sample is used to measure the selection efficiency, and includes radiative corrections.
Two processes are expected to contribute to photon emission: the inner bremsstrahlung 
photon emission, $\DKSee+\gamma_{IB}$; a two-photon decay with one virtual photon conversion, 
$\ks\to\gamma\gamma^*\to\gamma e^+ e^-$. The first process is simulated 
using the PHOTOS~\cite{Barberio} generator.
The events due to the second process are rejected by the kinematic cuts
used in the analysis.

\section{Data analysis}
\label{sec:ana}

The identification of \kl-interaction in the EMC is used to tag the presence
of \ks\ mesons. The mean decay lengths of \ks\ and \kl\ are 
$\lambda_S \sim 0.6\cm$ and $\lambda_L \sim 350\cm$, respectively. About $50\%$
of \kl's therefore reach the calorimeter before decaying. The \kl\ interaction 
in the calorimeter barrel (\kcr) is identified by requiring a cluster of energy above $125 \MeV$,
not associated with any track, and with a time 
corresponding to the \kl\ velocity in the $\phi$ rest frame, $\beta^\ast\sim 0.216$. 
Requiring $0.17 \le \beta^\ast \le 0.28$ we selected 
$\sim 650$ million \ks-tagged events (\kcr\ events in the following), which are used as a starting sample for the 
\DKSee\ search. 

As a first step of the signal search,  we select events with tracks of opposite 
charge having point of closest approach to the origin within a 
cylinder $4\cm$ in radius and $10\cm$ in length along the beam line. 
The two tracks are required to form a vertex with position in the transverse plane 
$\rho<4\cm$. Moreover, the track momenta ($p$) and polar angles ($\theta$) 
must satisfy the cuts: $120\MeV /c \le p \le 350\MeV /c$ and 
$30^{\circ} \le \theta \le 150^{\circ}$. The tracks must also reach the EMC
without spiralling, and have an associated cluster with energy $E_{cl} > 50 \MeV$ 
and position in the transverse plane $\rho_{cl}>60\cm$. These requirements ensure 
a redundant determination of the event-$T_0$ and allow us to evaluate the 
time of flight (TOF) for each particle.

\begin{figure}[h!]
  \begin{center}
    \includegraphics[totalheight=7.6cm]{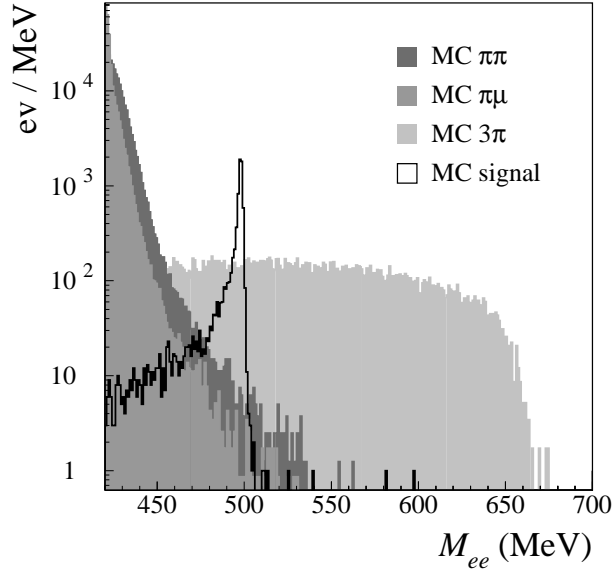}
    \caption{$\Mee$ MC spectra for signal (open histogram) and the main
background sources (gray histograms), as explained in the text; the signal corresponds to a 
BR of $1.6\times 10^{-4}$.}
    \label{fig:minvee}
  \end{center}
\end{figure}

The two-track invariant mass evaluated in the hypothesis of electron mass, $M_{ee}$, is used to 
reject the dominant background due to \DKSpippim.
We require $M_{ee}> 420 \MeV$, thus removing most of $\DKSpippim$ events which peak at $M_{ee}\sim 409 \MeV$,
with a resolution of $\sim 1\MeV$.
In order to reject tracks with a larger uncertainty on the fit parameters, 
we also require the propagated error on the  invariant mass, $\Delta M_{ee}$, to be less than $2.5\MeV$. 
In Fig.~\ref{fig:minvee}, the $M_{ee}$ distribution is shown for both MC signal and background. 
The background is due to the following sources: residual \DKSpippim\ events, populating the low 
$M_{ee}$ region, and \Dphipippimpio\ events, spreading over the whole spectrum. A \DKSpippim\ event
can have such a high value of $M_{ee}$ either because one track is badly reconstructed 
(\pipi\ component in the following) or because one pion decays to a muon before entering 
the DC and the vertex is reconstructed from a pion and
a muon track (\pimu\ component hereafter). 
A \Dphipippimpio\ event (\trepi\ component in the following) can satisfy the \ks\ tagging criteria 
from the presence of a machine background cluster (fake \kcr). 
At this stage of the analysis, we are left with $\sim 5\times10^5$ events. The efficiency for signal selection, 
given the \ks\ tagging, is $\sim 0.54$, as evaluated using MC. 

\begin{figure}[h!]
  \begin{center}    
    \includegraphics[totalheight=6.cm]{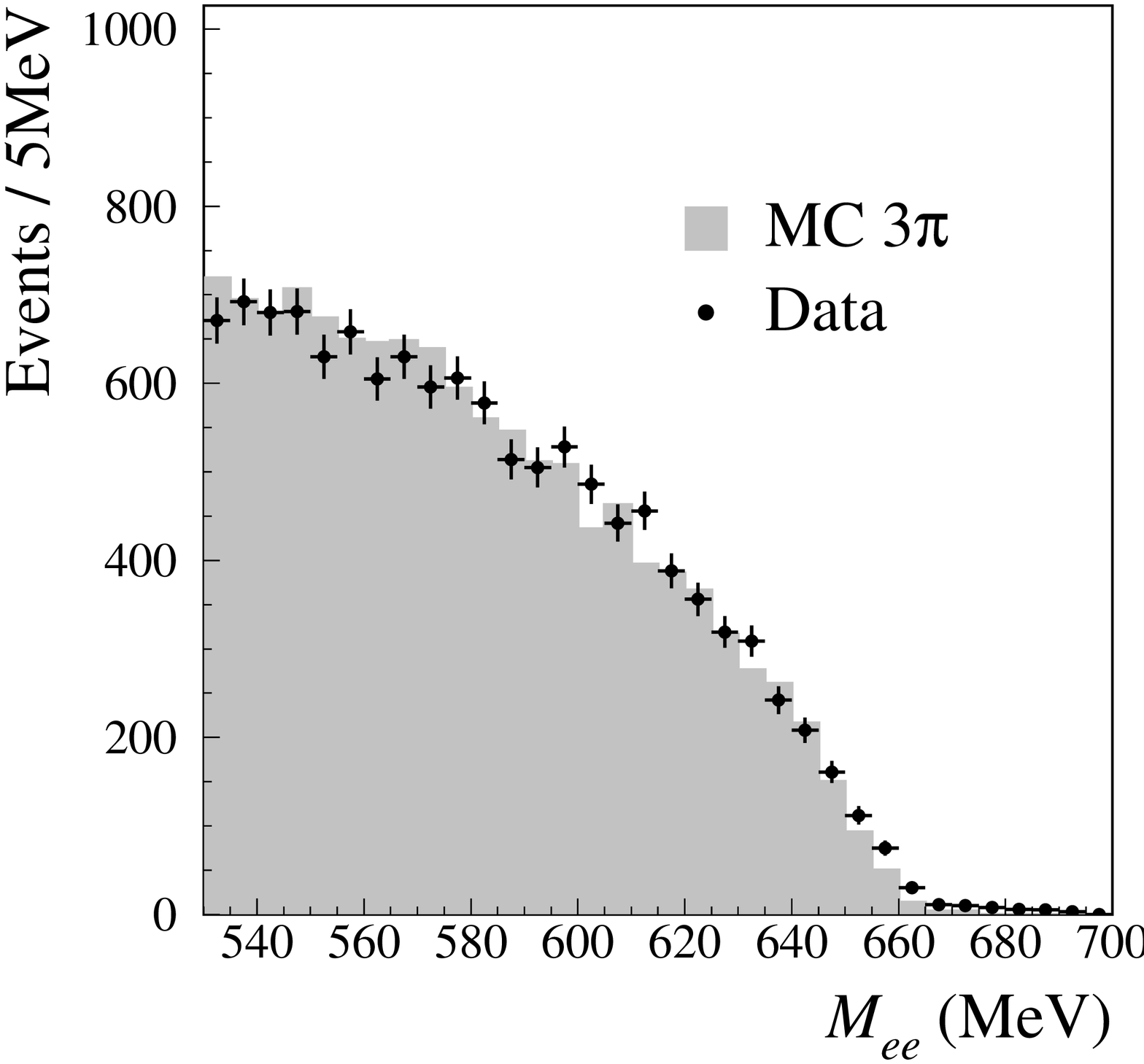}
    \includegraphics[totalheight=6.cm]{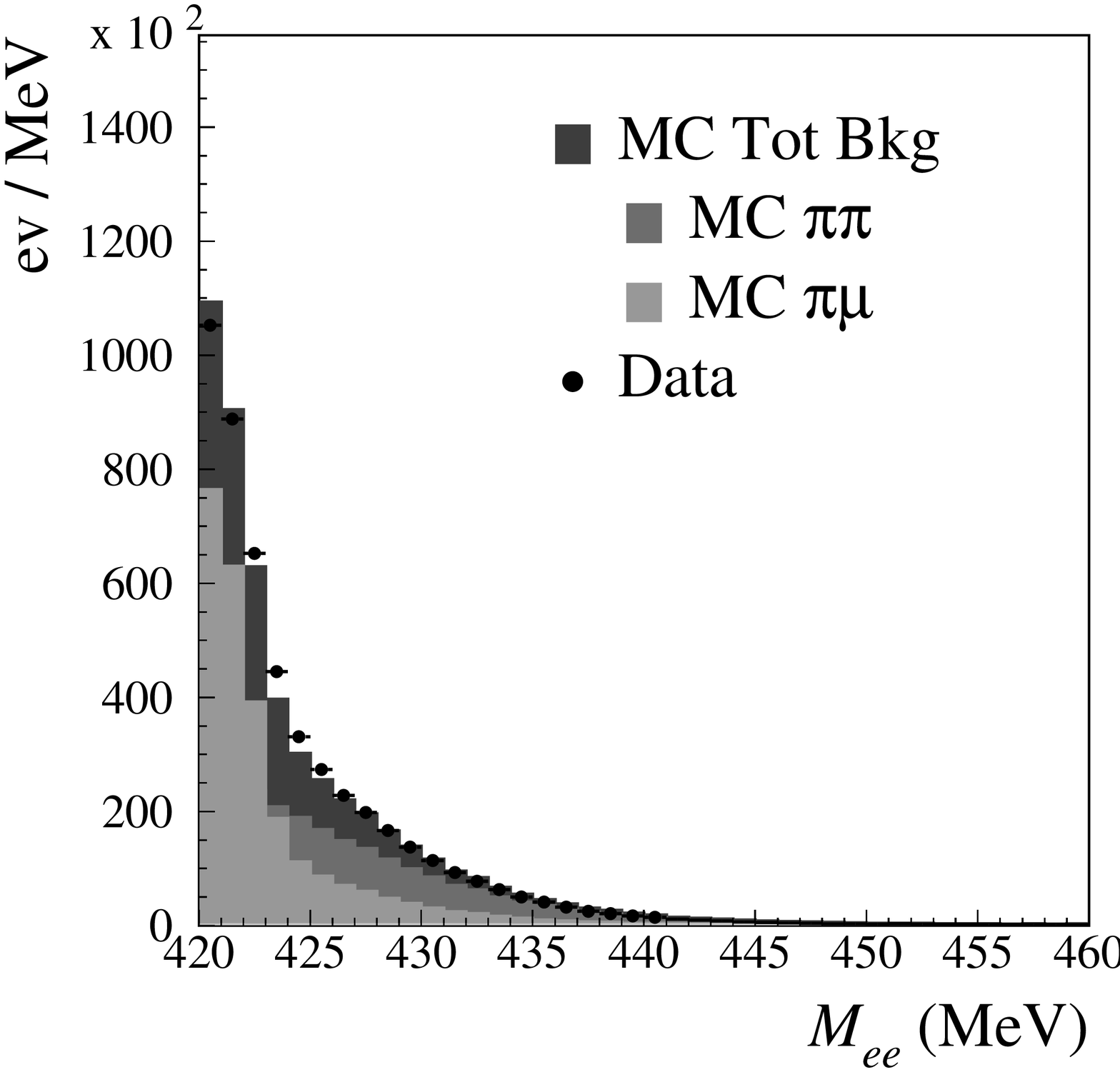}
    \caption{Data-MC comparison for $M_{ee}$ spectra in region 3 (left) and region 1 (right), after normalization;
      data are represented by the black points, MC background components by gray histograms.}
    \label{fig:sidebands}
  \end{center}
\end{figure}

The absolute background level is not taken directly from MC , but is obtained by normalization of data in the
region of signal sidebands.
The reliability of MC prediction is checked comparing with data after each step of the analysis. 
For this purpose, the $M_{ee}$ interval is divided into a signal region, around the kaon mass peak, 
and two sidebands:
\begin{equation}
\begin{array}{l}
 420 \le M_{ee}<460\MeV \qquad \textrm{region 1 (left sideband),}\\
 460 \le M_{ee}<530\MeV \qquad \textrm{region 2 (signal),}\\
 530 \le M_{ee}<700\MeV \qquad \textrm{region 3 (right sideband).}
\end{array}
\end{equation}
\pipi\ and \pimu\ background sources largely dominate on \trepi\ component in region 1, 
the opposite occurring in region 3. 
A scale factor for the \trepi\ component, $f_{\trepi}$, is therefore directly evaluated in region 3
as the ratio of the number of events in the data sample and the number of MC $3\pi$ events. 
We obtain:
\begin{equation}
\label{eq:f3pi}
f_{\trepi} = \frac{N({\rm data})}{N_{\trepi}({\rm MC})}=1.73\pm 0.03, 
\end{equation}
which has to be compared with a data/MC luminosity scale factor of $\simeq 0.86$.
The observed discrepancy is well understood, being due to the fact that MC underestimates the rate of
fake \kcr\ from machine background.
After normalization, the $M_{\mathrm{ee}}$ shape is well reproduced by MC \trepi\ sample, as shown 
in the left panel of Fig.~\ref{fig:sidebands}.
To obtain the scale factors $f_{\pipi}$ and $f_{\pimu}$ for the \pipi\ and \pimu\ components, we fit the \Mee\
distribution in region 1 to a linear combination of the MC background spectra with the \trepi\ component fixed as
in Eq.~\ref{eq:f3pi}.
The MC distribution after fit is compared to data in the right panel of Fig.~\ref{fig:sidebands}; 
we obtain:
\begin{equation}
 \begin{array}{l}
    f_{\pimu}  =  0.861\pm 0.005, \\  
    f_{\pipi}  =  1.249 \pm 0.008, \\
    \rho = -0.77,
  \end{array}
\end{equation}
where $\rho$ is the correlation factor and the errors quoted are statistical
only. The scale factors $f_{\pimu}$ and $f_{\pipi}$ have to be compared
with the expected data/MC ratio of $\simeq 0.73$.
A sizable deviation is observed for \pipi\ events,
which is expected since MC tends to underestimate the rate of events in the very far tails of
the tracking resolution.
After normalization, the number of background events, $N_{\mathrm{bkg}}^{\mathrm{MC}}$, is estimated as: 
\begin{equation}
N_{\mathrm{bkg}}^{\mathrm{MC}}= f_{\pipi}\times N_{\pipi}^{\mathrm{MC}} + f_{\pimu}\times N_{\pimu}^{\mathrm{MC}} + f_{\trepi}\times N_{\trepi}^{\mathrm{MC}}. 
\end{equation}

\begin{figure}[h!]
  \begin{center}    
    \includegraphics[totalheight=5.9cm]{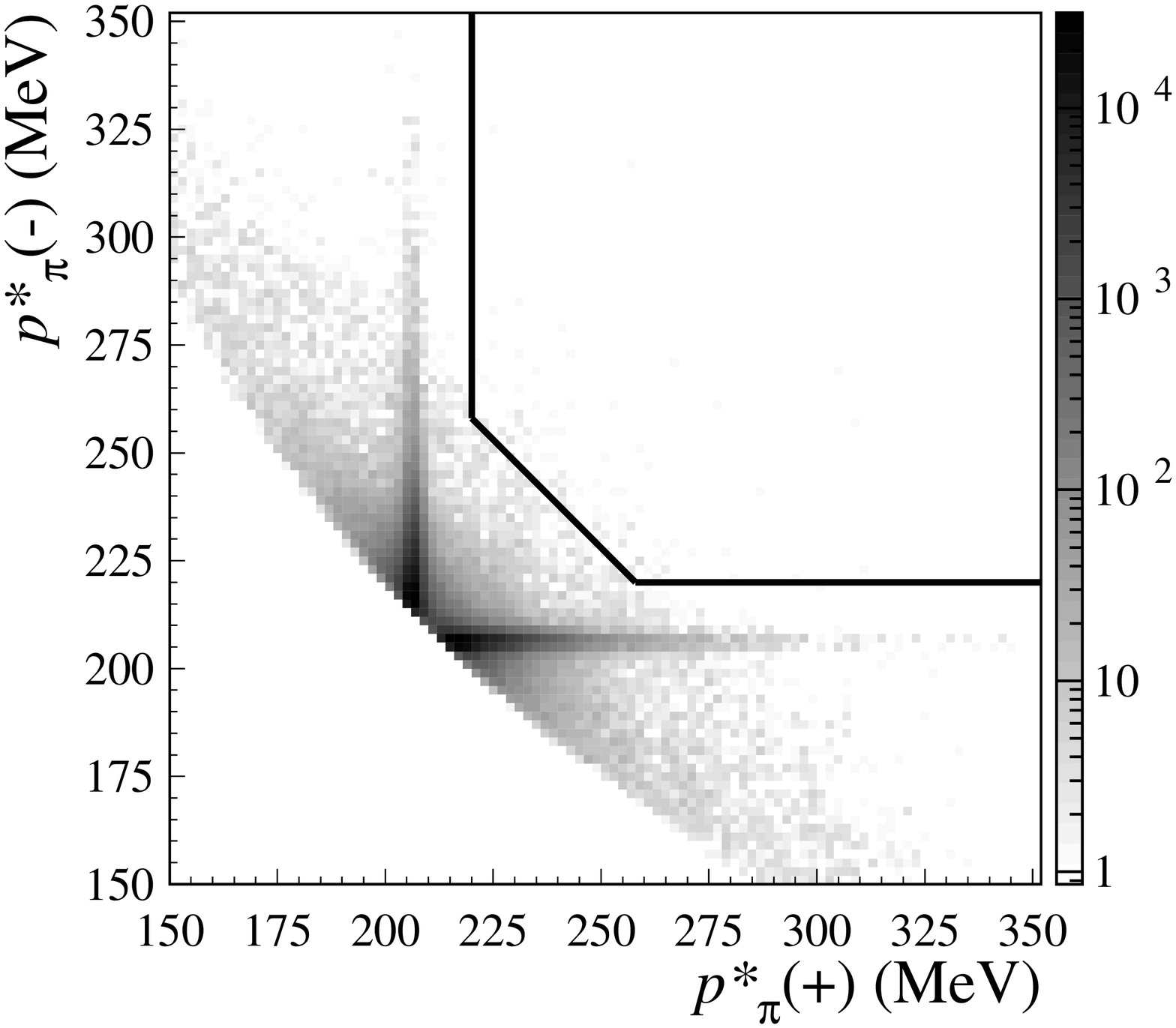}
    \includegraphics[totalheight=5.9cm]{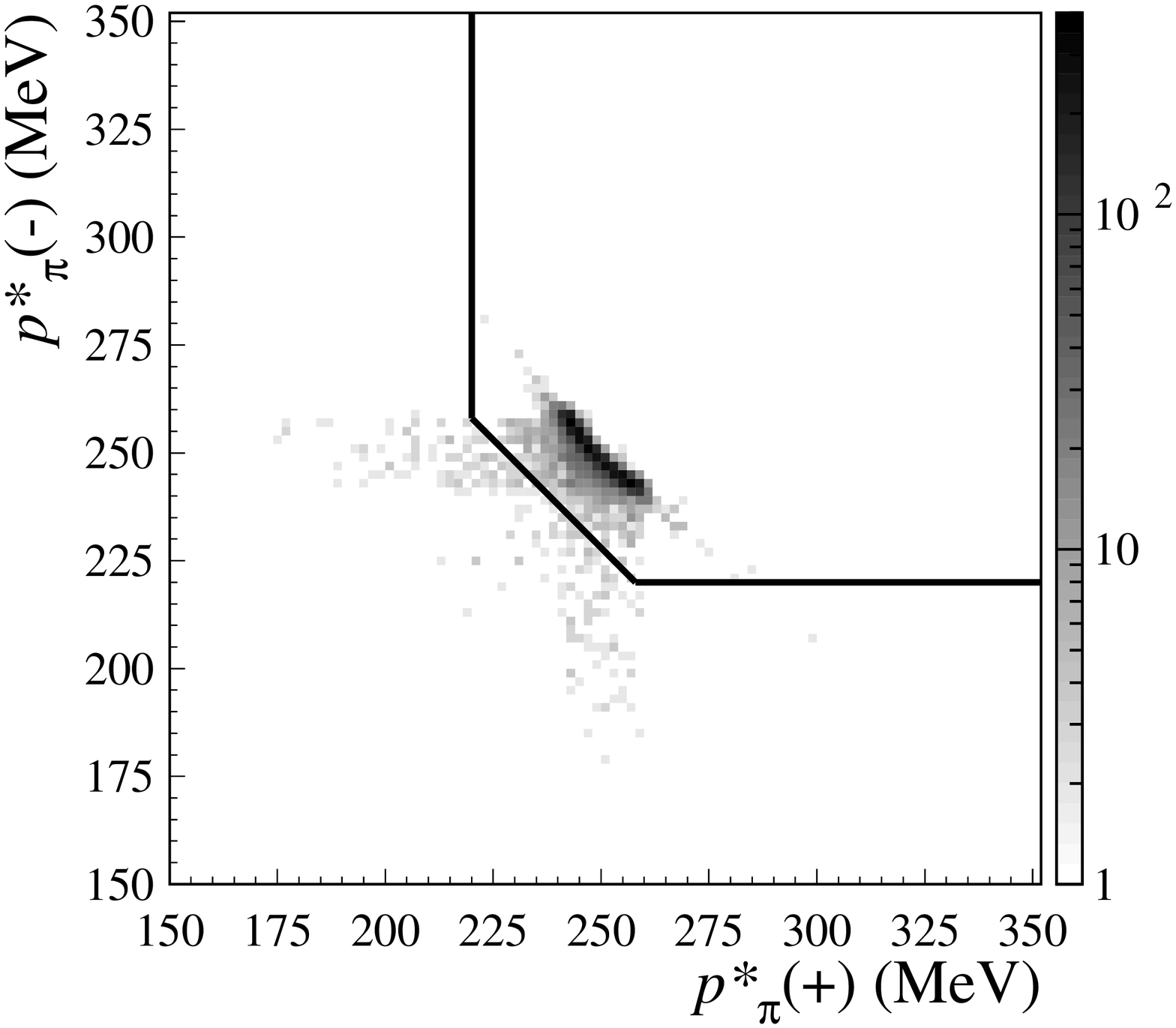}
    \caption{Scatter plot of track momenta $p^{\ast}_\pi(+)$ and $p^{\ast}_\pi(-)$ (assuming $\pi$ mass) 
             in the \ks\ rest frame for MC background (left) and MC signal (right); 
             selected region is shown by the solid line.}
    \label{fig:pstar}
  \end{center}
\end{figure}

Kinematics and topology can be further exploited to improve background rejection.
For most of \pipi\ and \pimu\ events, at least one pion track is well reconstructed, so that  
its momentum in the \ks\ rest frame $p_\pi^\ast$ peaks around 206\MeV, as expected for \DKSpippim\ decays. The
signal distribution populates higher values of $p_\pi^\ast$ (see Fig.~\ref{fig:pstar}). 
Therefore we require for both tracks $p^{\ast}_\pi>220\MeV /c$ and $p^{\ast}_\pi(+)+p^{\ast}_\pi(-)>478\MeV /c$, thus 
rejecting $\sim 99.9\%$ of \pipi\ and \pimu\ events and $\sim 8\%$ of signal events lying in 
the low \Mee\ tail.  

\begin{figure}[h!]
  \begin{center}    
    \includegraphics[totalheight=6.cm]{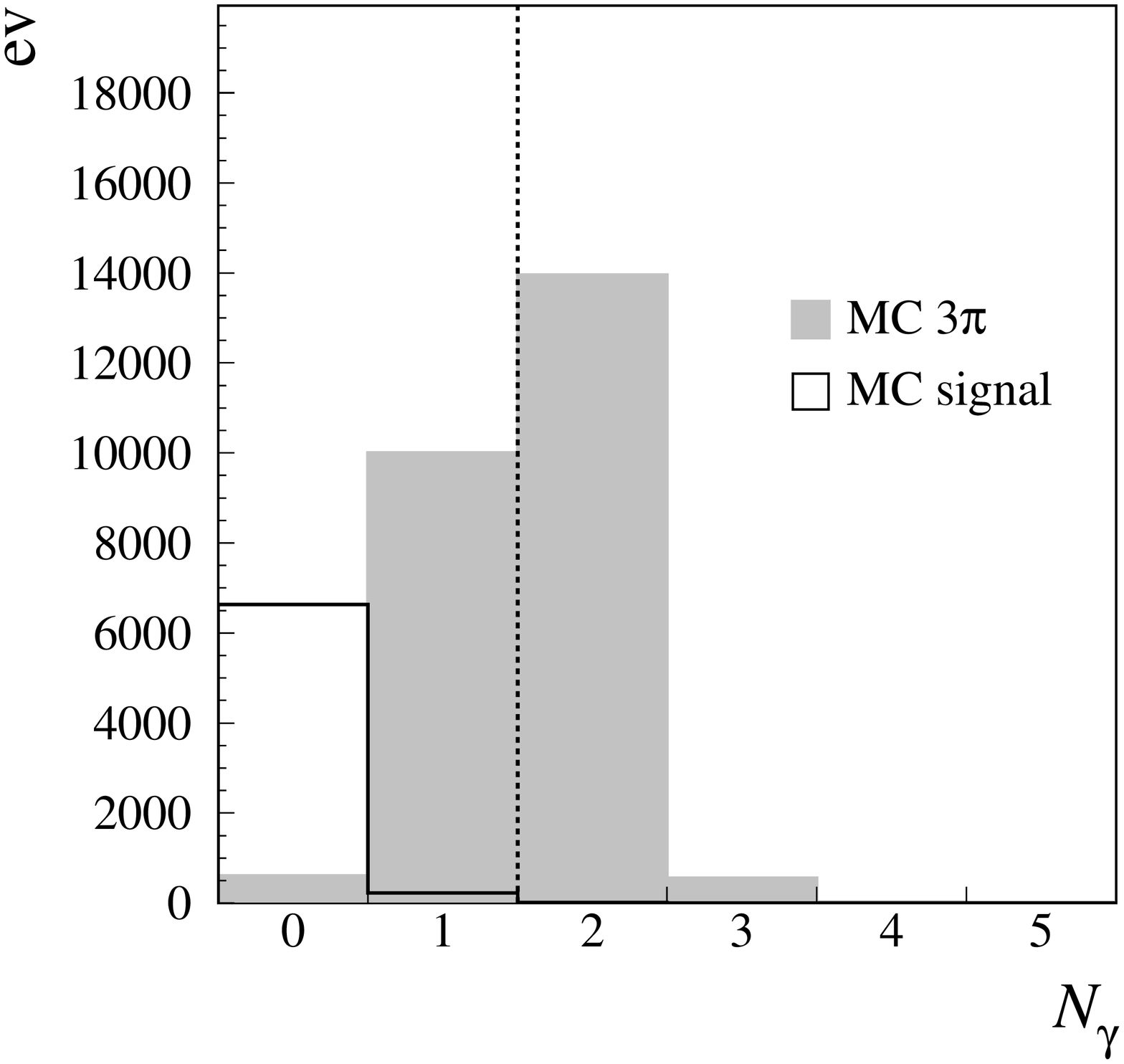}
    \includegraphics[totalheight=6.cm]{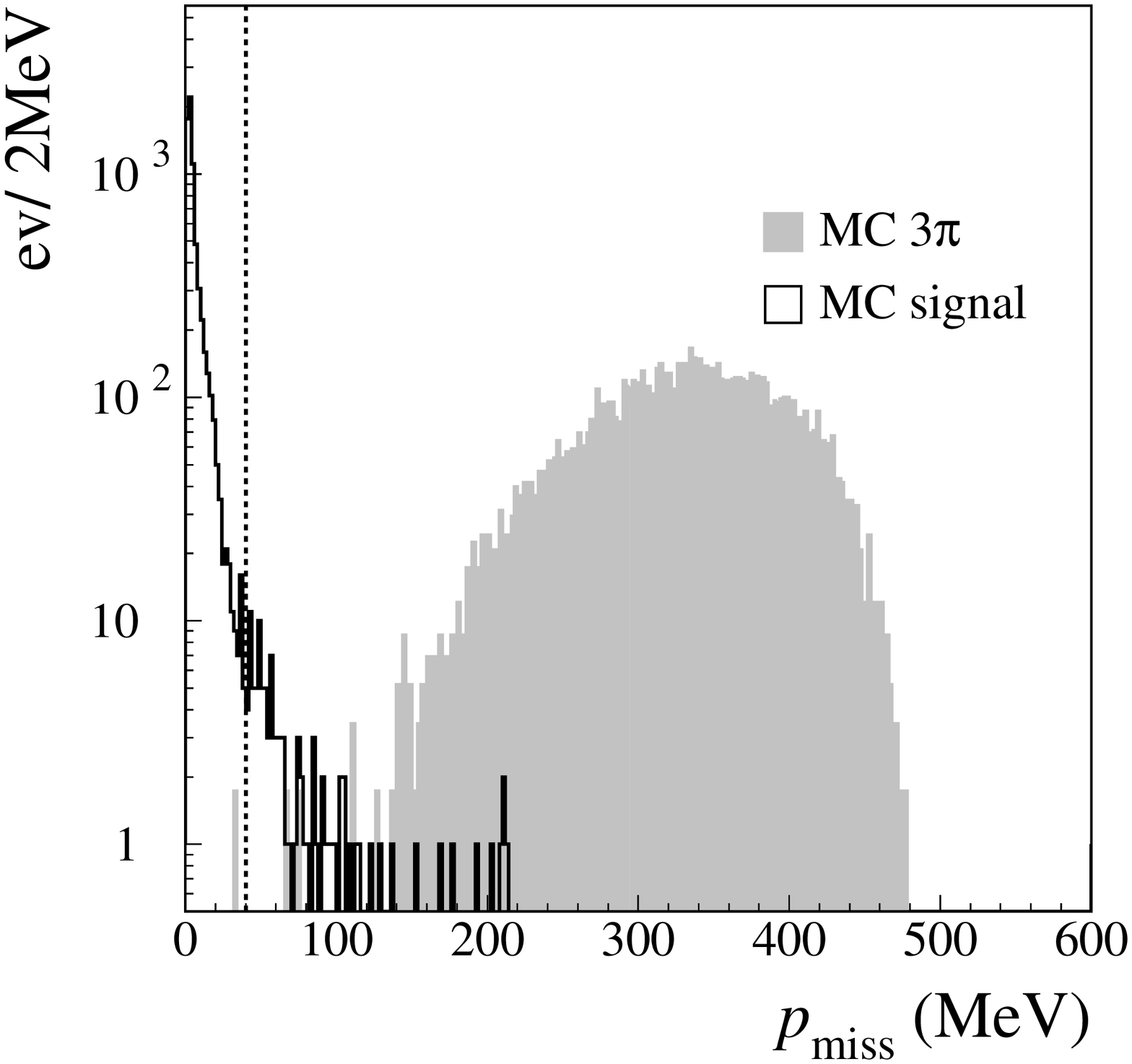}
    \caption{Distribution of $N_{\gamma}$ (left) and
      $\pmiss$ (right) for 
      MC signal (open histogram) and background events (gray histogram).}
    \label{fig:2cuts}
  \end{center}
\end{figure}

To further reject \trepi\ events we follow a two-step procedure.
First, we exploit the fact that in $\sim65\%$ of the cases the two photons 
originated in $\pi^0$ decay
are observed. Each $\gamma$ cluster is identified by TOF through the requirement 
$\delta t = t_{cl} - r_{cl}/c < 5\sigma_t$.
The number of detected $\gamma$'s is shown in the left panel of Fig.~\ref{fig:2cuts} for MC signal and 
\trepi\ events.
We reject events with $N_{\gamma}\ge2$, thus introducing a $\sim0.1\%$ loss of signal 
events.\footnote{Rejecting events with $N_{\gamma}=1$ would improve background rejection, 
  but would also introduce a systematic error in the evaluation of signal efficiency related 
  to photon radiation in the final state.}
Second, we cut on the total missing momentum of the $\phi$ decay, 
evaluated as $p_{\mathrm {miss}} = \vert  \vec{p}_\phi - \vec{p}_L - \vec{p}_S  \vert$, where
$\vec{p}_{L,S}$ are the neutral kaon momenta, and $\vec{p}_\phi$ is the $\phi$ momentum. 
The \ks\ momentum is evaluated from the charged track momenta, while the \kl\ momentum is measured from the \kcr\
cluster position and the \Pphi\ boost. The $p_{\mathrm {miss}}$ distribution is shown in the right panel 
of Fig.~\ref{fig:2cuts} for MC signal and \trepi\ events.
We reject events with $p_{\mathrm {miss}}\ge 40 \MeV$,
thus introducing a $\sim 2\%$ loss of signal events while reducing the \trepi\ background to 
a negligible level. 

A comparison between data and the expected background, $N_{\mathrm{bkg}}^{\mathrm{MC}}$, in the $M_{ee}$ 
sidebands after $p^\ast_\pi$ and $N_{\gamma}$ cuts is shown 
in Table~\ref{tab:sidebands}. This proves the reliability of the background simulation and of the normalization
procedure. 
\begin{table}[ht!]
\begin{center}
\begin{tabular}{|c|ccc|ccc| }
\hline
 & \multicolumn{3}{c|}{Region 1} & \multicolumn{3}{c|}{Region 3}\\
Cut & Data & $N_{\mathrm{bkg}}^{\mathrm{MC}}$& $\Delta/\sigma$ & Data  & $N_{\mathrm{bkg}}^{\mathrm{MC}}$& $\Delta/\sigma$\\ \hline  
$p^\ast_\pi>220\MeV$   &  3738 &  3980(100)   & -2.09  & 12107  & 12140(230)  & -0.11  \\ 
$N_{\gamma}<2$&        1516 &  1720(60)       & -2.83  & 5041   & 5090(120)   & -0.36  \\
\hline
\end{tabular}
\end{center}
\caption{Data counts and MC background estimates in the $M_{ee}$ sidebands after each cut; 
the difference $\Delta$ between data and MC in units of the total error is also reported.}
\label{tab:sidebands}
\end{table}
No event is observed in region 1, while one event remains in region 3 after $p_{\mathrm {miss}}$ cut,
for both data and MC background. The surviving MC event is a \trepi\ decay.

\begin{figure}[h!]
 \begin{center}    
    \includegraphics[totalheight=8.5cm]{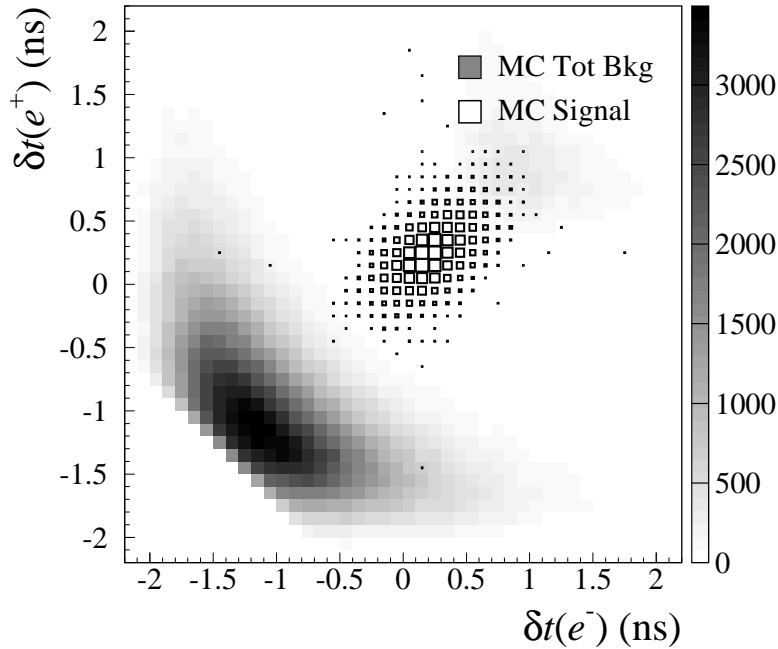}
   \caption{Scatter plot of $\delta t$ for the two charged tracks 
     for MC signal (box) and background events (gray), before background rejection cuts are applied.}
   \label{fig:deltat}
 \end{center}
\end{figure}

\begin{figure}[h!]
  \begin{center}    
    \includegraphics[totalheight=8.5cm]{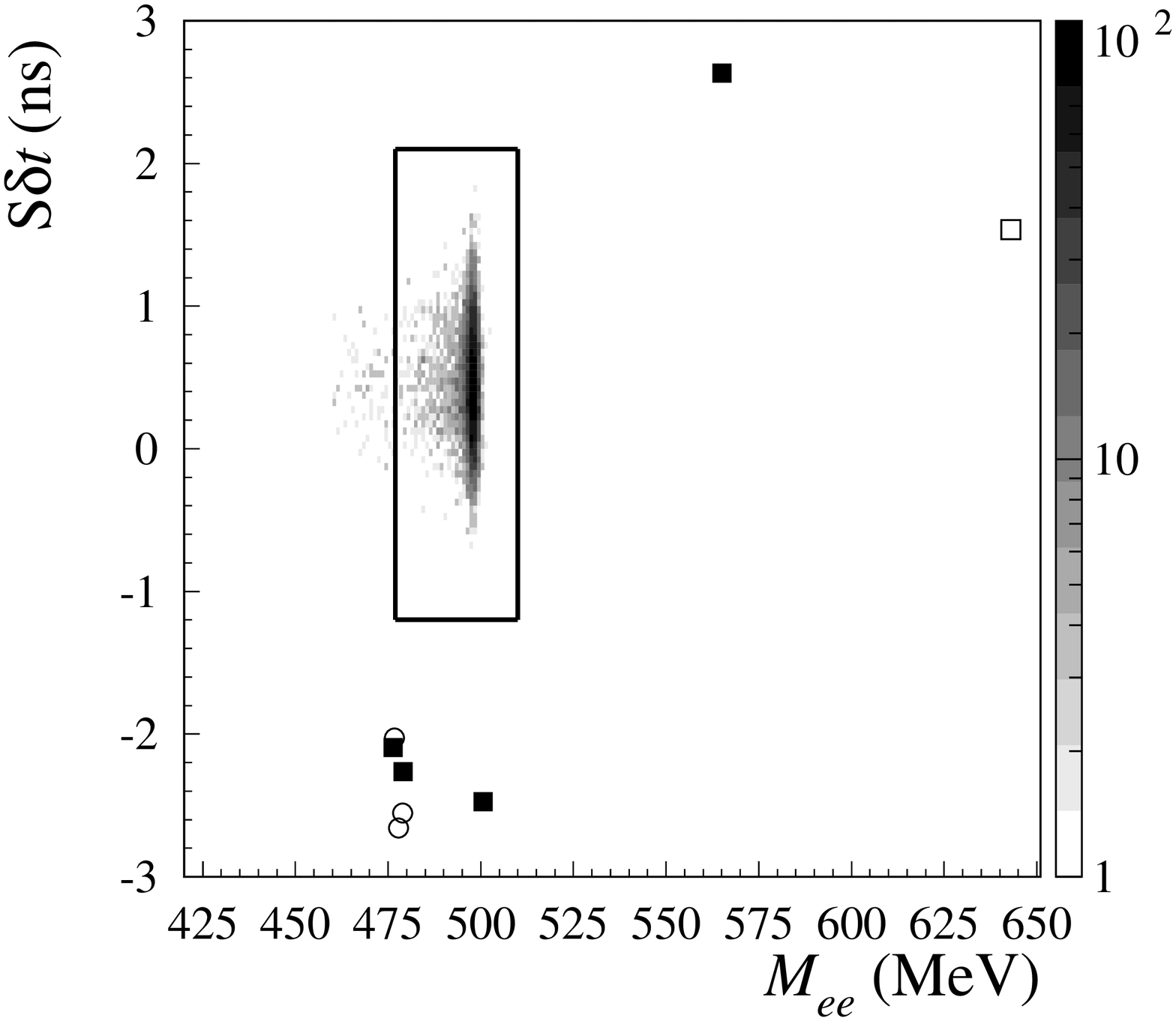}
    \caption{Scatter plot of $S\delta t$ versus $M_{ee}$ for data ($\blacksquare$), MC signal (gray),
             \DKSpippim\ ({\scriptsize $\bigcirc$}) and \Dphipippimpio\ ($\square$) 
after background rejection cuts.}
    \label{fig:sigbox}
  \end{center}
\end{figure}

At this stage of the analysis, in region 2 we count three events for data and three \pipi\ events for 
MC background.   
To improve background rejection, we exploit the particle identification capability
of the calorimeter. 
For this purpose, we evaluate the difference $\delta t(e) = t_{cl} - L/\beta(m_e) c$
between the measured cluster time and the expected particle time of flight under mass
hypothesis $m_e$, where $\beta(m_e) = p/(m_e^2+p^2)^{1/2}$.  
The high rejection capability provided by the TOF is demonstrated in Fig.~\ref{fig:deltat}, where the
scatter plot for $\delta t(e)$ of the two tracks is shown 
for MC signal and background events, before application of the $p^\ast_\pi$, $N_{\gamma}$ and $p_{\mathrm {miss}}$ cuts.
The best rejection is obtained by using the sum $S\delta t = \delta t(e^+) + \delta t(e^-)$. The
 scatter plot of $S\delta t$ as a function of \Mee\ is shown in 
Fig.~\ref{fig:sigbox} for all of the events surviving the background rejection cuts.
The signal box is then defined as follows:
\begin{equation}
 \begin{array}{rcl}
 477 < & M_{ee}& \le 510\, \MeV, \\
 -1.2 \le &S\delta t &\le 2.1\,  \mbox{ns}. 
 \end{array}
\end{equation}
This corresponds to a $\sim 9\sigma$ cut on $S\delta t$.
The lower bound of the \Mee\ range has been set to clearly define the fraction of 
IB spectrum which is accepted in our selection:
all \DKSee\ events with a radiated photon with energy above $20\MeV$ are rejected.
This cut reduces to a negligible level the other contribution to radiative decay,
$\ks\to\gamma\gamma^*\to\gamma e^+ e^-$, which is strongly peaked for $\Mee \sim 2 m_{e}$ . 
Following Ref.~\cite{Bergstrom}, we evaluated 
$BR(\ks\to\gamma\gamma^*\to\gamma e^+ e^-, E_\gamma^*<20\MeV)\sim 6\times 10^{-12}$, 
which is far beyond our experimental sensitivity.
After the signal box cut we evaluate a signal efficiency given the tag 
$\epsilon_{\mathrm{sig}}(\mathrm{sele}) = 0.479(6)$.

Applying the signal box selection, we observe no event on data. 
Equally, no MC background event falls in the signal box, see Fig.~\ref{fig:sigbox}.
In the conservative assumption of no background, we obtain the upper limit  
on the expected number of signal events $\UL(N_{\mathrm{sig}}) = 2.3$, at $90\%$ CL.

\section{Systematic uncertainties}
\label{sec:syst}
Since no background subtraction has been made, there is no need to asses
any systematic error on the scale factors applied to the MC.
The selection efficiency for \DKSee\ has been corrected to take account of small differences 
between data and MC tracking efficiency. 
The latter has been evaluated on a \Dphipippimpio control sample, both for data and MC.
A systematic uncertainty of $0.9\%$ has been evaluated by varying the correction in its
allowed range. 
In order to evaluate the systematic uncertainty related to the cuts on $\Delta M_{ee}$, 
$N_{\gamma}$ and $p_{\mathrm {miss}}$, 
we have used a \DKSpippim\ control sample, 
selected by a tight cut around the kaon mass for both data and MC events.
The systematics on the previous cuts has been then evaluated on the control sample
as the difference between data and MC efficiencies for each of the above requirements, 
as listed in Table~\ref{tab:syssum}.
Finally, we checked the effect of IB photon emission on the selection efficiency.
The presence of radiated photons affects indeed the shape of the momentum distribution, and thus
the efficiency of the $p_{\pi}^*$ selection.
A systematic uncertainty of  $0.8\%$ has been evaluated by comparing results obtained with or without the
inclusion of photon radiation in the final state.
\begin{table}[h!]
\begin{center}
\begin{tabular}{c|c}
  Source & Fractional error  \\
\hline
Tracking & 0.9\%   \\
$\Delta M_{ee}$ & 1.4\% \\
$N_{\gamma}$    & 0.5\%  \\
$p_{\mathrm {miss}}$     & 1.3\%  \\
$p^*_{\pi}$     & 0.8\%     \\
\hline
total  & 2.3\%      \\
\end{tabular}
\end{center}
\caption{Summary of systematic uncertainties.}
\label{tab:syssum}
\end{table}
All of the contributions to the systematic uncertainty are listed in
Table~\ref{tab:syssum}.
The total fractional error is $2.3\%$. 

\section{Upper limit evaluation}

The upper limit on BR(\DKSee) is evaluated by normalizing  $\UL(N_{\mathrm{sig}})$ to the number of 
\DKSpippim\ events, $N_{\pippim}$, counted in the same sample of \ks\ tagged events:
\begin{eqnarray}
 \UL\left (BR(\DKSee)\right )& = & \nonumber \\
 \UL(N_{\mathrm{sig}})     &\times & 
\frac{\epsilon_{\pippim}(\kcr )}{
\epsilon_{\mathrm{sig}}(\kcr ) } \times
\frac{\epsilon_{\pippim}(\mathrm{sele} )}{
\epsilon_{\mathrm{sig}}(\mathrm{sele} ) } \times
\frac{BR(\DKSpippim)}{N_{\pippim}}, \nonumber 
\end{eqnarray}
where $\epsilon(\kcr)$ and $\epsilon(\mathrm{sele})$ are the tagging efficiency and the selection
efficiency, given \kcr\ tag, for each channel.
\DKSpippim\ events are identified by requiring the presence of two tracks of opposite 
charge, selected with the same cuts as for \DKSee, with no additional requirements on
invariant mass, kinematical quantities, and particle identification.
The selection efficiency for both channels is evaluated from MC, with corrections 
extracted from data control samples.
We obtain $\epsilon_{\pippim}(\mathrm{sele}) = 0.6102(5)$ and $\epsilon_{\mathrm{sig}}(\mathrm{sele}) = 0.479(6)$.
The ratio of tagging efficiencies slightly differs from unity, 
$\epsilon_{\pippim}(\kcr)/\epsilon_{\mathrm{sig}}(\kcr)= 0.9634(1)$. This 
dependence of the tagging efficiency on the \ks\ decay mode is due to 
a small difference in the determination of the event-$T_0$  
in presence of electrons or pions in the final state, which affects 
the measurement of the \kl\ velocity. This bias is evaluated from data
using \DKSpippim\ and \DKSpiopio\ events~\cite{kspipi}.
Using $N_{\pippim} = 217\,422\,768$ and BR(\DKSpippim) from Ref.~\cite{pdg06},
we obtain:
\begin{equation}
\UL\left (\mathrm{BR} \left (\DKSee \left (\gamma
\right)\right)\right) =  9 \times 10^{-9}, \; {\rm at} \;90\%\,{\rm CL}.
\end{equation}
The effect of systematic uncertainty (see Sec.~\ref{sec:syst}) on the BR evaluation 
is accounted for by a Gaussian smearing of the total efficiency in the UL calculation. 
Our measurement improves by a factor of $\sim16$ on the
CPLEAR result~\cite{cplearksee}, for the first time including radiative
corrections in the evaluation of the upper limit.

\section*{Acknowledgments}
We thank the DA$\Phi$NE team for their efforts in maintaining low background running
conditions and their collaboration during all data-taking.
We want to thank our technical staff:
G.F. Fortugno and F. Sborzacchi for ensuring the efficient operation of the KLOE 
computing facilities; M.~Anelli for his continuous attention to the gas system 
and the safety of the detector;
A.~Balla, M.~Gatta, G.~Corradi and G.~Papalino for maintenance of the electronics;
M.~Santoni, G.~Paoluzzi and R.~Rosellini for the general support to the detector;
C.~Piscitelli for his help during major maintenance periods.
This work was supported in part by EURODAPHNE, contract FMRX-CT98-0169;
by the German Federal Ministry of Education and Research (BMBF) contract 06-KA-957;
by the German Research Foundation (DFG), 'Emmy Noether Programme', contracts DE839/1-4;
by INTAS, contracts 96-624, 99-37; and by the EU Integrated Infrastructure Initiative 
HadronPhysics Project under contract number RII3-CT-2004-506078.

\bibliographystyle{elsart-num}

\end{document}